# An efficient surrogate-aided importance sampling framework for reliability analysis

Wang-Sheng LIU [1], Wen-Jun CAO [2,3], and Sai Hung CHEUNG [1*]

[1] *School of Civil & Environmental Engineering, Nanyang Technological University, Block N1, 50 Nanyang Avenue, Singapore 639798*

[2] *Department of Civil and Environmental Engineering, National University of Singapore, Block E1A, No.1 Engineering Drive 2, Singapore 117576*

[3] *ETH Zurich, Future Cities Laboratory, Singapore-ETH Centre, Singapore 138602*

**Abstract:** Surrogates in lieu of expensive-to-evaluate performance functions can accelerate the reliability analysis greatly. This paper proposes a new two-stage framework for surrogate-aided reliability analysis named Surrogates for Importance Sampling (S4IS). In the first stage, a coarse surrogate is built to gain the information about failure regions; the second stage zooms into the important regions and improves the accuracy of the failure probability estimator by adaptively selecting support points therein. The learning functions are proposed to guide the selection of support points such that the exploration and exploitation can be dynamically balanced. As a generic framework, S4IS has the potential to incorporate different types of surrogates (Gaussian Processes, Support Vector Machines, Neural Network, etc.). The effectiveness and efficiency of S4IS is validated by five illustrative examples, which involve system reliability, highly nonlinear limit-state function, small failure probability and moderately high dimensionality. The implementation of S4IS is made available to download at (the link will be added once the paper is accepted for publication).

# 1 Introduction

(Background: reliability analysis) Reliability analysis is very essential for ensuring a system to perform as designed when subjected to different sources of uncertainties. In the context of engineering, the measure of the reliability is the failure probability. Given a model whose uncertain parameters are represented by a vector of $d$ random variables $\boldsymbol{\Theta} = [\Theta_1, \cdots, \Theta_d]^T$, the failure probability $P_F$ for a specific failure event $F$ can be written as:

$$P_F = \int \mathbb{I}_F(\boldsymbol{\theta}) p_0(\boldsymbol{\theta}) d\boldsymbol{\theta} = \mathbb{E}_{p_0}[\mathbb{I}_F(\boldsymbol{\theta})] \tag{1}$$

where $\boldsymbol{\theta} = [\theta_1, \cdots, \theta_d]^T \in \Omega \subset \mathbb{R}^d$ denotes the parameter value of $\boldsymbol{\Theta}$ in the parameter space $\Omega$; $\mathbb{I}_F(\cdot)$ is the binary indicator function of the failure event $F$ which equals to one if $\boldsymbol{\theta}$ belongs to the failure region $\Omega_F$ and zero otherwise; $p_0(\cdot)$ is the joint probability density function (PDF) of $\boldsymbol{\Theta}$; and $\mathbb{E}_{p_0}[\cdot]$ denotes the expectation for $\boldsymbol{\Theta} \sim p_0(\boldsymbol{\theta})$. The failure region $\Omega_F$ is defined by the performance function $g(\boldsymbol{\theta})$, i.e., $\Omega_F = \{\boldsymbol{\theta}: g(\boldsymbol{\theta}) \leq 0\}$. The corresponding limit-state function $g(\boldsymbol{\theta}) = 0$ divides the parameter space $\Omega$ into the failure and safety regions.

(Reliability methods) A considerable amount of literature has been focusing on developing efficient and accurate ways to calculate the failure probability. One widely used set of reliability methods is based on the linear or quadratic approximation of the limit-state function around the most probable point (MPP) using Taylor series approximation, usually known as the first or second order reliability methods (FORM or SORM), respectively [1, 2]. But the error brought by the approximation tends to be large when the limit-state function is highly nonlinear and/or there exists multiple MPPs in the failure region; furthermore, the computational cost in the search of MPP increases drastically with the number of random variables. Another set of reliability methods is based on stochastic sampling. Among them, Monte Carlo Simulation (MCS) has been acknowledged as a robust technique where the failure probability is approximated statistically through stochastic sampling. In its crude form, MCS first simulates $N$ samples from the PDF $p_0(\boldsymbol{\theta})$; then evaluates the corresponding values of the performance function $g(\boldsymbol{\theta})$ and counts the number of failure samples $N_F$; finally, an estimator $\hat{P}_F$ can be given by $N_F/N$. Though $\hat{P}_F$ is asymptotically unbiased, the convergence rate of the crude MCS is low. For example, to estimate a small failure probability of $10^{-t}$, it takes a very large size of $10^{t+2}$ samples for the estimator to achieve a Coefficient of Variation (*CoV*) close to 10%. Advanced stochastic simulation techniques have been developed to reduce the required sample size, including Subset Simulation (SS) [3], Importance Sampling (IS) [4], and Spherical Subset Simulation (S³) [5]. Despite the remarkable improvement in efficiency, they still require a moderately large number of samples to be evaluated (e.g., in SS, the sample size in each level is typically set to 1000).

(Motivation for surrogates) In practice, the performance function $g(\boldsymbol{\theta})$ of a complex system often involves implicit and time-demanding numerical models such as finite element models which are widely used by engineers to predict the responses of a system. What's worse, in the attempt to better mimic the behavior of a system, numerical models tend to be more and more sophisticated, making the model evaluation increasingly computationally expensive. Consequently, the aforementioned reliability methods are likely to be inapplicable for complex systems when the computation resource is limited.

(Review of surrogates and learning strategies) To alleviate this problem, surrogates (also referred to as metamodels or response surfaces) can be built in lieu of the original expensive performance functions. In the literature, different surrogates have been employed to perform reliability analysis, such as polynomials [6, 7], Gaussian process (GP or Kriging) [8-12], polynomial chaos expansion (PCE) [13, 14], artificially neural networks (ANN) [15, 16] and support vector machines (SVM) [17]. The most critical part of efficiently building a surrogate is probably the adaptive infill strategy, also referred to as adaptive design of experiment or refinement scheme in some literature. Instead of densely filling the parameter space with support points, an adaptive infill strategy often starts with a coarse surrogate built upon a small number of support points, and then refines it progressively using judiciously selected support points. Many GP-based adaptive infill strategies have been reported in the last decade in conjunction with stochastic-sampling-based reliability methods. For example, Bichon et al. [9] presented an Efficient Global Reliability Analysis (EGRA) method where an Expected Feasibility Function (EFF) was maximized by using the global optimization algorithm to select the best next infill support points in each iteration. More recently, Echard et al. [18] proposed a new method AK-MCS which combines GP with crude Monte Carlo Simulation. In this method, an active learning function called U function was developed to guide the selection of sequential support points. Later, a modified version AK-IS [12] was proposed using the same learning function, which resorts to importance sampling for solving small-failure-probability cases. Alternatively, Dubourg et al. [11] adaptively built a GP surrogate as a substitute of the performance function and used it to construct the quasi optimal importance sampling density.

Surrogate-aided reliability analysis is very parsimonious with the help of progressive refinement, but most adaptive infill strategies in the current literature are customized for GP and specific reliability analysis problems only. The paper presents a new framework for surrogate-aided reliability analysis named Surrogates for Importance Sampling (S4IS). This framework is robust to different types of reliability analysis problems (e.g., high dimensionality and small failure probability) and has the potential to incorporate different types of surrogates. The estimation of the failure probability and the building of the surrogate are performed in two stages. In Stage 1, support points are generated to build a coarse surrogate which can identify multiple failure regions; the next stage focuses more on the important regions to improve the failure probability estimator. The adaptive infill of support points in two stages is under the guidance of two different learning functions such that the dynamic balance between the exploration and exploitation can be achieved. The proposed learning functions are the convex combination of several objectives.

The remainder of this paper is organized as follows. Section 2 introduces the importance sampling technique for reliability analysis as well as the existing learning functions used to guide the infill of support points. In Section 3, the proposed two-stage framework S4IS is discussed in detail. To validate the proposed framework, five illustrative examples are investigated in Section 4, which involves system reliability problem, highly nonlinear limit-state function, different reliability levels and moderately high dimensionality. The performance of S4IS is compared with the crude MCS, FORM and AK-IS.

## 2 Surrogate-aided importance sampling

### 2.1 Importance sampling

It is common that the system failure $F$ in Equation (1) is a rare event, meaning the failure region $\varOmega_F$ is in the tail part of $p_0(\boldsymbol{\theta})$. In this case, the crude MCS may cannot afford to simulate sufficient failure samples in order to achieve a small variance of the estimated $P_F$. Alternatively, Importance Sampling (IS) [4] is widely used as a variance reduction technique which computes the expectation with respect to a proposed instrumental PDF.

Given an instrumental PDF $q(\boldsymbol{\theta})$ supported on $\Omega \subset \mathbb{R}^d$, the failure probability in Equation (1) can now be rewritten as:

$$P_{F,IS} = \int \mathbb{1}_F(\boldsymbol{\theta}) \frac{p_0(\boldsymbol{\theta})}{q(\boldsymbol{\theta})} q(\boldsymbol{\theta}) \mathrm{d}\boldsymbol{\theta} = \mathbb{E}_q\left[\frac{\mathbb{1}_F(\boldsymbol{\theta}) p_0(\boldsymbol{\theta})}{q(\boldsymbol{\theta})}\right] \tag{2}$$

Instead of sampling from the original PDF $p_0(\boldsymbol{\theta})$, we can sample from $q(\boldsymbol{\theta})$ and adjust $\mathbb{1}_F(\boldsymbol{\theta})$ with the likelihood ratio $p_0(\boldsymbol{\theta})/q(\boldsymbol{\theta})$. Now the IS estimator $\hat{P}_{F,IS}$ can be expressed as:

$$\hat{P}_{F,IS} = \frac{1}{N_{IS}} \sum_{i=1}^{N_{IS}} \mathbb{1}_F(\boldsymbol{\theta}^{(i)}) \frac{p_0(\boldsymbol{\theta}^{(i)})}{q(\boldsymbol{\theta}^{(i)})}, \boldsymbol{\theta}^{(i)} \sim q(\boldsymbol{\theta}) \tag{3}$$

where $\{\boldsymbol{\theta}^{(i)}; i: 1 \to N_{IS}\}$ denotes a set of $N_{IS}$ i.i.d (independent and identical distributed) samples from $q(\boldsymbol{\theta})$. The IS estimator is unbiased and its variance is estimated by:

$$\begin{aligned} Var[\hat{P}_{F,IS}] &= \frac{1}{N_{IS}(N_{IS}-1)} \sum_{i=1}^{N_{IS}} \left[\frac{\mathbb{1}_F(\boldsymbol{\theta}^{(i)}) p_0(\boldsymbol{\theta}^{(i)})}{q(\boldsymbol{\theta}^{(i)})} - \hat{P}_{F,IS}\right]^2 \\ &= \frac{1}{N_{IS}-1}\left[\frac{1}{N_{IS}}\sum_{i=1}^{N_{IS}} \frac{\mathbb{1}_F(\boldsymbol{\theta}^{(i)}) p_0(\boldsymbol{\theta}^{(i)})^2}{q(\boldsymbol{\theta}^{(i)})^2} - \hat{P}_{F,IS}^2\right] \end{aligned} \tag{4}$$

The selection of the instrumental PDF is of crucial importance in reducing the variance. It is not difficult to derive that minimizing the variance gives the optimal instrumental PDF

$$q^*(\boldsymbol{\theta}) = \frac{\mathbb{1}_F(\boldsymbol{\theta}) p_0(\boldsymbol{\theta})}{\int \mathbb{1}_F(\boldsymbol{\theta}) p_0(\boldsymbol{\theta}) \mathrm{d}\boldsymbol{\theta}} = \frac{\mathbb{1}_F(\boldsymbol{\theta}) p_0(\boldsymbol{\theta})}{P_F} \tag{5}$$

Although this zero-variance optimal density in Equation (5) is not usable without knowing $P_F$, it can guide the selection procedure of $q(\boldsymbol{\theta})$ and provide insights to the design of the importance sampling scheme. In practice, quasi-optimal instrumental PDFs are commonly used.

## 2.2    Building the surrogate adaptively

(Surrogate for what) Assume the performance function of interest $g(\boldsymbol{\theta})$ is a time-demanding black-box function. The main goal is to build a surrogate $\hat{g}(\boldsymbol{\theta})$ to the original function from a dataset of support points $\mathcal{S} = \{(\boldsymbol{\theta}^{(i)}, y^{(i)}); i: 1 \to N_s\}$ so that all failure regions $\Omega^F$ over the parameter space $\Omega$ can be identified accurately. In other words, in contrast with general-purpose surrogates that are optimized to be accurate over the whole parameter space, surrogates for reliability analysis focus on the accuracy of failure region boundaries $\partial \Omega^F$ defined by the limit-state function $g(\boldsymbol{\theta}) = 0$. Given the surrogate $\hat{g}(\boldsymbol{\theta})$, the binary indicator function in Equation (2) and (3) can be approximated by:

$$\hat{\mathbb{1}}_F(\boldsymbol{\theta}) = \begin{cases} 1, & \hat{g}(\boldsymbol{\theta}) \leq 0 \\ 0, & \hat{g}(\boldsymbol{\theta}) > 0 \end{cases} \tag{6}$$

(Active and passive infill strategy) An infill strategy should be employed to select a finite number of support points aiming at gaining the maximal information for building the surrogate. Instead of using dense support points for fitting a general-purpose surrogate, an adaptive infill strategy starts with a small size of support points to fit an initial surrogate and then new support points are sequentially selected to refine the previous surrogate. This way is proved to be far more efficient than the non-adaptive one since it is customized to perform the analysis of interest (e.g., reliability analysis in this paper).

(How to build adaptively) Finding the best next support point(s) in each iteration is guided by some criteria or objectives in terms of optimization. In general, on one hand, we expect the new support point to explore the under-explored regions where existing support points are sparse; one the other hand, the previously explored regions should be exploited to improve the accuracy over the regions of interest. These two criteria are referred to as *exploration* and *exploitation*, respectively. In the context of reliability analysis, exploration involves adding more support points to sparsely populated regions in order to identify the failure region(s) and exploration involves zooming into the regions that contribute most to the integral for estimating the failure probability. As exploration and exploitation are conflicting goals, the problem of finding the next support point(s) is essentially a multi-objective optimization problem.

Various *Learning functions* have been proposed as objectives to solve the multi-objective optimization problem. An early learning function called Expected Feasibility Function (EFF) was introduced by Bichon et al. [9] in the Efficient Global Reliability Analysis (EGRA). EFF converts a multi-objective problem to a single objective problem, acting as an indicator of how well the true value is expected to satisfy the limit-state function $g(\boldsymbol{\theta}) = 0$ in its vicinity of $\pm \epsilon$

$$EFF(\boldsymbol{\theta}) = \int_{-\epsilon}^{+\epsilon} (\epsilon(\boldsymbol{\theta}) - |\hat{g}(\boldsymbol{\theta})|) f_{\hat{G}} \mathrm{d}\hat{g}(\boldsymbol{\theta}) \tag{7}$$

where $\hat{g}(\boldsymbol{\theta})$ is a realization of a Gaussian Process (GP) distributed as $f_{\hat{G}}$; and $\epsilon$ is set to be proportional to the standard deviation of $\hat{g}(\boldsymbol{\theta})$ (e.g., $\epsilon = 2\sigma_{\hat{g}}$ in [9]). The EFF in Equation (7) can be expressed in the analytical form when random variables $\boldsymbol{\Theta}$ are transformed into the uncorrelated standard normal space via isoprobabilistic transformations [19-21]. The support point with the maximum expected feasibility is found by a global optimization method. It can be observed from Equation (7) that the optimization of EFF favors points that are close to the limit-state boundary (small $|g(\boldsymbol{\theta})|$) and have large variance in the prediction (large $\sigma_{\hat{g}}$)

Similarly, another learning function in terms of the prediction mean and variance of a GP surrogate, called U-function, was introduced in AK-MCS [18] and AK-IS [12]

$$U(\boldsymbol{\theta}) = \frac{|\mu_{\hat{G}}(\boldsymbol{\theta})|}{\sigma_{\hat{G}}(\boldsymbol{\theta})} \tag{8}$$

where $\mu_{\hat{G}}(\boldsymbol{\theta})$ and $\sigma_{\hat{G}}(\boldsymbol{\theta})$ denote the prediction mean and standard deviation function of GP, respectively. The next point(s) can be obtained by minimizing $U(\boldsymbol{\theta})$. A sampling-based approach is used to replace the global optimization step in EGRA, where the values of $U(\boldsymbol{\theta})$ are calculated for a large number of candidate support points drawn from the original distribution of random variables.

The above two learning functions interpreted exploitation as the closeness to limit-state boundary but neglected the fact that the high-density regions in the failure domain contribute more to the failure probability than the low-density ones. To address this issue, Sun et al. [22] combined the statistical information provided by the GP surrogate with the joint PDF of random variables by proposing a new learning function named Least Expected Improvement Function (LIF).

## 3 The proposed framework

In the existing framework described in [12] and [23], the first stage consists in the search of the most probable point (MPP) and reliability analysis using FORM; in the second stage, IS is performed by proposing a multivariate Gaussian distribution as the instrumental PDF, which is centered on the MPP and with covariance being the identity matrix. However, this framework can only incorporate Gaussian Process (GP) as the surrogate; furthermore, it is not able to identify multiple MPPs or multiple failure regions.

In this paper, a new two-stage framework for surrogate-aided importance sampling called Surrogates for Importance Sampling (S4IS) is presented as illustrated in Figure 1. In the first stage, a coarse surrogate is built adaptively to identify all failure regions in the parameter space. The second stage starts with the construction of the Gaussian Mixture (GM) instrument PDF for IS based on the identified failure regions, then proceeds to the refinement of the surrogate mainly over the important regions to reduce the error of the IS estimator. Two different learning functions proposed in this framework are formulated by the convex combination of different exploration and exploitation objectives and are applicable to different types of surrogates. Also, the learning functions in the two stages balance the exploration and exploitation dynamically, with the one in the second stage focusing more on exploitation to ensure the fast convergence. As shown in Figure 1, infill support points in Stage 1 scatter more uniformly over the whole parameter space than those in Stage 2.

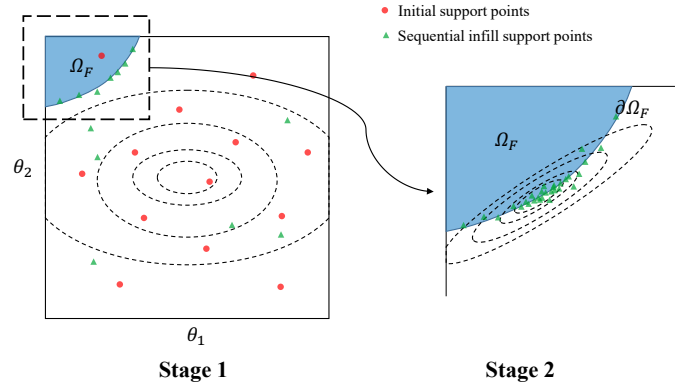

**Figure 1: Schematic diagram of the proposed two-stage framework Surrogate for Importance sampling (S4IS)**

## 3.1 Stage 1

### 3.1.1 Initialization

Throughout this paper, an isoprobabilistic transformation $T$ [19-21] is applied to the original random variables $\mathbf{\Theta}$ to transform them into independent standard normal random variables $\mathbf{U}$, i.e., $\mathbf{U} = T(\mathbf{\Theta})$. The IS estimator in Equation (3) can be rewritten as:

$$\hat{P}_{F,IS} = \frac{1}{N_{IS}} \sum_{i=1}^{N_{IS}} \mathbb{I}_F\big(\mathbf{u}^{(i)}\big) \frac{p_n\big(\mathbf{u}^{(i)}\big)}{q\big(\mathbf{u}^{(i)}\big)}, \mathbf{u}^{(i)} \sim q(\mathbf{u}) \qquad (9)$$

where $p_n(\cdot)$ is the multivariate standard normal PDF and the failure region in **U**-space can be given by $\Omega_F = \{\mathbf{u}: g(T^{-1}(\mathbf{u})) \leq 0\}$.

To explore the parameter space, a large number $N_{c1}$ (in this paper, $N_{c1} = \min(10^4, 10^d)$) of samples are randomly generated from a distribution $q_1(\mathbf{u})$ in the first stage. They are treated as the inputs of candidate support points $\mathbf{u}_{c1} = \{\mathbf{u}^{(i)} \sim q_1(\mathbf{u}); i: 1 \to N_{c1}\}$. After selecting a small subset $\mathbf{u}_{s1,0}$ from $\mathbf{u}_{c1}$, the outputs of the subset are evaluated

$$y_{s1,0} = \{y^{(i)} = g\left(T^{-1}(\mathbf{u}^{(i)})\right); \mathbf{u}^{(i)} \in \mathbf{u}_{s1,0}\} \tag{10}$$

where $y_{s1,0}$ denote the outputs of the time-demanding performance function for $\mathbf{u}_{s1,0}$. Given the dataset of initial support points $\mathcal{S}_{1,0} = \{(\boldsymbol{\theta}_{s1,0}, y_{s1,0})\} = \{(T^{-1}(\mathbf{u}_{s1,0}), y_{s1,0})\}$, the initial surrogate $\hat{g}_{1,0}$ can be built to predict the output

$$\hat{y} = \hat{g}_{1,0}(\boldsymbol{\theta}) = \hat{g}_{1,0}(T^{-1}(\mathbf{u})) \tag{11}$$

In this paper, we suggest the size of $\mathbf{u}_{s1,0}$ to be $N_{s1,0} = \max[12, (d+1)(d+2)/2]$ where $d$ is the dimension of $\mathbf{u}$.

The inputs of candidate support points $\mathbf{u}_{c1}$ in the first stage can be simulated from the standard normal distribution, i.e., $q_1(\mathbf{u}) = p_n(\mathbf{u})$. However, for cases of small failure probabilities, only a few or even not a single $\mathbf{u}_{c1}$ will scatter in the failure regions, leading to at least two problems: (i) since the inputs of support points is selected from the candidates, the surrogate fitted using the corresponding support points will hardly be accurate near the limit-state boundary; (ii) multiple failure regions cannot be effectively identified. In this paper, $\mathbf{u}_{c1}$ are generated uniformly in the hypercubic space ranging from -5 to 5 in each dimension so as to explore the whole parameter space, i.e., $q_1(\mathbf{u}) = \frac{1}{10^d}, \mathbf{u} \in [-5,5]^d$.

### 3.1.2 Adaptive refinement of the surrogate

At the beginning of iteration $k$ ($k = 1,2 \cdots$), the current surrogate $\hat{g}_{1,k}$ has been fitted using the dataset of support points $\mathcal{S}_{1,k} = \{(\boldsymbol{\theta}_{s1,k}, y_{s1,k})\} = \{(T^{-1}(\mathbf{u}_{s1,k}), y_{s1,k})\}$ (for $k = 1$, $\hat{g}_{1,k} = \hat{g}_{1,0}$ and $\mathcal{S}_{1,k} = \mathcal{S}_{1,0}$). Our aim is to find the best next input of support point from the candidates $\mathbf{u}_{c1} \setminus \mathbf{u}_{s1,k}$. The existing inputs of support points $\mathbf{u}_{s1,k}$ is excluded from the candidate set $\mathbf{u}_{c1}$ to avoid duplication.

The learning function is essential to the selection of the location of new support points. For the first stage, the learning function achieves the balance between exploration and exploitation by minimizing the distance to the limit-state boundary (exploration) and meanwhile maximizing the distance to the existing support points (exploitation)

$$LF_1(\mathbf{u}) = |\hat{g}_{1,k}(T^{-1}(\mathbf{u}))| - \|\mathbf{u}, \mathbf{u}_{s1,k}\|_{min} \qquad (12)$$

where $\|\mathbf{u}, \mathbf{u}_{s1,k}\|_{min}$ is the minimum distance between $\mathbf{u}$ and $\mathbf{u}_{s1,k}$ in iteration $k$. Therefore, the new input of support point selected from $\mathbf{u}_{c1}\setminus\mathbf{u}_{s1,k}$ can be obtained as:

$$\mathbf{u}_k^* = \underset{\mathbf{u}^{(i)}\in\mathbf{u}_{c1}\setminus\mathbf{u}_{s1,k}}{\mathrm{argmin}} \left(LF_1(\mathbf{u}^{(i)})\right) \qquad (13)$$

Equation (12) differs from the learning functions in Equation (7) and (8) in a way that the exploration is conducted by maximizing the minimal distance between the new and existing inputs of support points rather than maximizing the prediction variance. Since the learning function in Equation (12) does not rely on the prediction statistics provided by GP, it can be applied to other types of surrogates (SVM, ANN, PCE, etc.). Also, it converts the multi-objective optimization problem into a single objective one by equally weighted sum of objectives. The convex combination of objectives ensures that the optimal solution is a Pareto optimal.

Various distance metrics can be used to measure the distance between two points $\mathbf{u}^{(i)}$ and $\mathbf{u}^{(j)}$ in real-valued vector spaces. The Euclidean distance is suggested in this paper, which can be calculated by:

$$\|\mathbf{u}^{(i)}, \mathbf{u}^{(j)}\| = \sqrt{\sum_{t=1}^{d} \left(\mathrm{u}_t^{(i)} - \mathrm{u}_t^{(j)}\right)^2} \qquad (14)$$

Therefore, the minimal distance $\|\mathbf{u}, \mathbf{u}_{s1,k}\|_{min}$ where $\mathbf{u}_{s1,k}$ is a set of inputs of support points can be given by:

$$\|\mathbf{u}, \mathbf{u}_{s1,k}\|_{min} = \underset{\mathbf{u}^{(i)}\in\mathbf{u}_{s1,k}}{\mathrm{argmin}} \left(\|\mathbf{u}, \mathbf{u}^{(i)}\|\right) \qquad (15)$$

The selected input $\mathbf{u}_k^*$ in Equation (13) is evaluated by the implicit performance function to obtain the output $y_k^* = g(T^{-1}(\mathbf{u}_k^*))$. For the next iteration, the dataset of support points is updated as $\mathcal{S}_{1,k+1} = \mathcal{S}_{1,k} \cup \{(T^{-1}(\mathbf{u}_k^*), y_k^*)\}$ and the corresponding updated surrogated is built as $\hat{g}_{1,k+1}$ accordingly.

### 3.1.3 Stopping criteria

In each iteration $k$ ($k = 1,2 \cdots$), the failure probability is estimated by:

$$\hat{P}_{F1,k} = \frac{1}{N_{c1}} \sum_{i=1}^{N_{c1}} \hat{\mathbb{I}}_F\left(T^{-1}(\mathbf{u}^{(i)})\right) \frac{p_n(\mathbf{u}^{(i)})}{q_1(\mathbf{u})}, \mathbf{u}^{(i)} \in \mathbf{u}_{c1} \qquad (16)$$

where the indicator function based on the surrogate $\hat{g}_k$ is written as:

$$\hat{\mathbb{I}}_F\left(T^{-1}(\mathbf{u})\right) = \begin{cases} 1, & \hat{g}_{1,k}(T^{-1}(\mathbf{u})) \leq 0 \\ 0, & \hat{g}_{1,k}(T^{-1}(\mathbf{u})) > 0 \end{cases} \qquad (17)$$

Note that $\hat{P}_{F1,k}$ may not be a good estimator because $p_n(\cdot)$ and $q_1(\cdot)$ can have different supports, but it can be treated as a measure of how well all candidates $\mathbf{u}_{c1}$ for the first stage have been classified into the safety and failure regions.

To decide in which iteration to terminate the Stage 1, early stopping criterion is used in this paper. We stop in iteration $n_{it1}$, if

$$\frac{\left|\hat{P}_{F1,n_{it1}} - \frac{1}{a_1}\sum_{k=n_{it1}+1-a_1}^{n_{it1}} \hat{P}_{F1,k}\right|}{\frac{1}{a_1}\sum_{k=n_{it1}+1-a_1}^{n_{it1}} \hat{P}_{F1,k}} \leq \varepsilon_1 \qquad (18)$$

where $a_1$ is a positive integer and $\varepsilon_1$ is the tolerance to the relative error of the estimated failure probability with respect to the mean failure probabilities from previous $a_1$ iterations. The setting $a_1 = 5$ works well for the proposed framework. And $\varepsilon_1$ controls the accuracy of the built surrogate and is set to 0.01. Another stopping criterion is $n_{it1}$ exceeding the maximal allowable number of iterations $[n_{it1}]$ for Stage 1, i.e., $n_{it1} > [n_{it1}]$.

At the end of the first stage, we can have (i) a coarse estimator of the failure probability $\hat{P}_{F1,n_{it1}}$; (ii) a coarse surrogate $\hat{g}_{1,n_{it1}}$; (iii) failure samples which are generated by evaluating all candidates $\mathbf{u}_{c1}$ using $\hat{g}_{1,n_{it1}}$. The results can be further utilized in the second stage to improve the estimation of the failure probability as we will discuss in the next section. It is also interesting to find that Stage 1 degrades to AK-MCS [18] if $q_1(\mathbf{u}) = p_n(\mathbf{u})$ in Equation (16) but with a different learning function.

### 3.1.4 High dimensionality

Sufficient inputs of candidate support points should be generated to ensure an effective exploration. Considerable attention must be paid to the input size $N_{c1}$ when the dimension $d$ is large ($d \geq 10$). Specifically, for the space defined by $[-5,5]^d$, an input size $10^d$ is required to achieve a density of one point per unit volume. Even though these inputs of candidates are evaluated only by the surrogate which requires relatively small computing efforts, the computational burden is likely to be prohibitive when $d$ becomes large for non-high-performance-computer users.

An alternative solution for exploration is to combine the search of multiple most probable points (MPP) with FORM as discussed in [1]. Unlike the sampling-based exploration, [1] searches for multiple MPPs via optimization by taking advantage of the first order gradient information of the performance function, making it very efficient even in high dimensional space. As a result, MPPs as well as a FORM estimator of the failure probability can be obtained. The support points generated during the optimization procedure are used to fit the initial surrogate for Stage 2. This exploration solution is suggested when $d$ is large and computers with large memory and computing power are not available.

## 3.2 Stage 2

### 3.2.1 Instrumental probability density function

Similar to [10], in this paper we use Gaussian Mixture (GM) as the instrumental PDF $q_2(\mathbf{u})$ in Stage 2 in order to take multiple MPPs into account. The GM is made up of $K$ equally weighted Gaussian distributions centered on the MPPs $\mathbf{u}_{MPP}$

$$q_2(\mathbf{u}) = \frac{1}{K} \sum_{t=1}^{K} \mathcal{N}(\mathbf{u}|\mathbf{u}_{MPP,t}, \mathbf{1}) \tag{19}$$

where the mean $\mathbf{u}_{MPP,t}$ is the MPP in the cluster $t$ and $\mathbf{1}$ represents the identity matrix. The MPPs $\mathbf{u}_{MPP}$ are determined approximately based on the failure samples from Stage 1 which are denoted as $\mathbf{u}_{c1,F} = \{\mathbf{u}^{(i)} \in \mathbf{u}_{c1}; \hat{g}_{1,n_{it1}}(T^{-1}(\mathbf{u}^{(i)})) \leq 0\}$ (or based on [1]). Specifically, first $K$-means algorithm [24] is used to partition $\mathbf{u}_{c1,F}$ into $K$ clusters. In each cluster $t$, the failure sample with the largest $p_n(\mathbf{u})$ is selected as the MPP for this cluster, i.e.,

$$\mathbf{u}_{MPP,t} = \underset{\mathbf{u}^{(i)} \in \mathbf{u}_{c1,F}^{(t)}}{\operatorname{argmax}} \left( p_n(\mathbf{u}^{(i)}) \right) \tag{20}$$

where $\mathbf{u}_{c1,F}^{(t)}$ is the set of failure samples in cluster $t$. It should be noted that the number of clusters $K$ may not equal to the number of failure regions. But it is not a problem as shown in the examples if we set a reasonably large $K$. The reason is that $K$ being larger than the number of failure regions simply leads to partitioning the connected failure regions and samples simulated from the GM $q_2(\mathbf{u})$ will still scatter over all the failure regions.

For initialization in Stage 2, the initial surrogate is $\hat{g}_{2,0} = \hat{g}_{1,n_{it1}}$ and the initial dataset of support points $\mathcal{S}_{2,0} = \{(\boldsymbol{\theta}_{s1,n_{it1}}, y_{s1,n_{it1}})\} = \{(T^{-1}(\mathbf{u}_{s1,n_{it1}}), y_{s1,n_{it1}})\}$. To calculate the IS estimator and further refine the surrogate, a large number of samples $\mathbf{u}_{c2} = \{\mathbf{u}^{(i)} \sim q_2(\mathbf{u}); i: 1 \to N_{c2}\}$ are generated from $q_2(\mathbf{u})$ and evaluated by $\hat{g}_{2,0}$. Then, the initial IS estimator in the second stage can be given by:

$$\hat{P}_{F2,0} = \frac{1}{N_{c2}} \sum_{i=1}^{N_{c2}} \hat{\mathbb{I}}_F \left( T^{-1}(\mathbf{u}^{(i)}) \right) \frac{p_n(\mathbf{u}^{(i)})}{q_2(\mathbf{u})}, \mathbf{u}^{(i)} \in \mathbf{u}_{c2} \tag{21}$$

Also, the samples $\mathbf{u}_{c2}$ serves as the inputs of candidate support points in Stage 2 whose outputs will be evaluated by the original performance function if selected. The sample size of $\mathbf{u}_{c2}$ is large in general (e.g., $N_{c2} = 10000$).

### 3.2.2 Adaptive refinement of the surrogate

Differing from the learning function $LF_1(\cdot)$ for the first stage, the learning function in the second stage puts more emphasis on the exploitation task by zooming into the important regions. Observing Equation (21), we can find that in the failure regions the larger the likelihood ratio $p_n/q_2$ is, the more it contributes to the IS estimator. By taking this into account, the learning function in Stage 2 adds another exploitation term to $LF_1(\cdot)$, which can be written as:

$$LF_2(\mathbf{u}) = |\hat{g}_{2,k}(T^{-1}(\mathbf{u}))| - \|\mathbf{u}, \mathbf{u}_{s1,k}\|_{min} - \log\left(\frac{p_n(\mathbf{u})}{q_2(\mathbf{u})}\right) \tag{22}$$

Like $LF_1(\cdot)$, the multi-objective optimization is converted to a single objective optimization problem by weighted convex combination of objectives.

In iteration $k$ ($k = 1,2 \cdots$), the current surrogate $\hat{g}_{2,k}$ has been fitted using the dataset of support points $\mathcal{S}_{2,k} = \{(\boldsymbol{\theta}_{s2,k}, y_{s2,k})\} = \{(T^{-1}(\mathbf{u}_{s2,k}), y_{s2,k})\}$ (for $k = 1$, $\hat{g}_{2,k} = \hat{g}_{2,0}$ and

$S_{2,k} = S_{2,0}$). The new input of support points can be selected from $\mathbf{u}_{c2}\setminus\mathbf{u}_{s2,k}$ by minimizing the learning function

$$\mathbf{u}_k^* = \underset{\mathbf{u}^{(i)}\in\mathbf{u}_{c2}\setminus\mathbf{u}_{s2,k}}{\mathrm{argmin}} \left(LF_2(\mathbf{u}^{(i)})\right) \tag{23}$$

Again, $\mathbf{u}_k^*$ is evaluated by the original performance function. New support point $(T^{-1}(\mathbf{u}_k^*), y_k^*)$ is appended to $S_{2,k}$ to update the surrogate.

### 3.2.3 Stopping criteria

Similar to the first stage, the iterative procedure terminates when either the IS estimator converges or the number of iterations $n_{it2}$ exceeds the predefined maximal allowable number of iterations $[n_{it2}]$. The convergence criterion is expressed as:

$$\frac{\left|\hat{P}_{F2,n_{it1}} - \frac{1}{a_2}\sum_{k=n_{it2}+1-a_2}^{n_{it2}} \hat{P}_{F2,k}\right|}{\frac{1}{a_2}\sum_{k=n_{it2}+1-a_2}^{n_{it2}} \hat{P}_{F2,k}} \leq \varepsilon_2 \tag{24}$$

where $a_2 = 5$ and $\varepsilon_2 = 0.001$.

### 3.3 Flowchart

The implementation of S4IS is summarized in Figure 2. To begin the process, a large number of inputs of candidate support points (or candidates) are generated from $q_1(\cdot)$ for Stage 1, which will later be selected adaptively to refine the surrogate in each iteration. To explore the whole parameter space effectively, heavy-tailed distributions $q_1(\cdot)$ such as uniform distributions are suggested. Based on the information about the failure regions obtained in Stage 1, Stage 2 starts with the construction of the instrumental probability density function $q_2(\cdot)$ for IS. Again, after selecting inputs of support points from the candidates generated from $q_2(\cdot)$, they are evaluated by the performance function to further refine the surrogate. Balancing dynamically between exploration and exploitation is achieved in the two stages by (i) different candidates to be evaluated by the performance function; (ii) different learning functions to select the best inputs of support points.

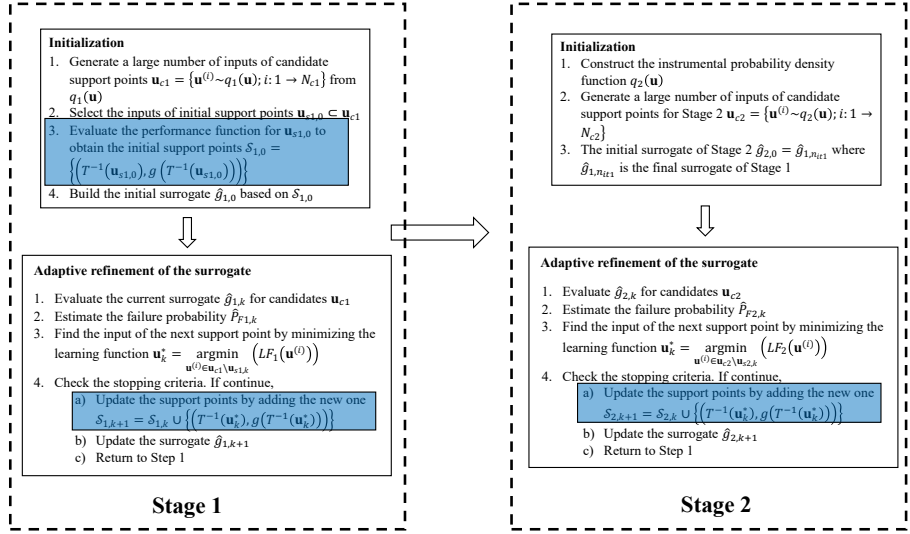

Figure 2: Flowchart of the proposed framework S4IS. Filled rectangles represent where the performance function is evaluated.

## 4 Illustrative examples

In this section, five examples are discussed to illustrate the effectiveness and efficiency of S4IS for different types of reliability analysis problems, which are featured by the system reliability, the involvement of nonlinear limit-state functions, small failure probability, and moderately high dimensionality. The efficiency is measured in terms of $N_{eval}$, i.e., the number of performance function evaluations required to achieve a small Coefficient of Variation ($CoV$). For all examples below, we use Gaussian Process (GP) as the surrogate and ten replicates are run to compute the statistics of the final estimator of the failure probability.

### 4.1 Example 1: a series system with four failure regions [18, 25]

This example involves a series system, in which the performance function returns the smallest value of four component performance function. The purpose of this example is to validate the proposed framework for system reliability analysis and multiple failure regions. The performance function is:

$$g(\boldsymbol{\theta}) = \min \begin{bmatrix} 3 + 0.1(\theta_1 - \theta_2)^2 - \dfrac{\sqrt{2}(\theta_1 + \theta_2)}{2} \\ 3 + 0.1(\theta_1 - \theta_2)^2 + \dfrac{\sqrt{2}(\theta_1 + \theta_2)}{2} \\ (\theta_1 - \theta_2) + 3\sqrt{2} \\ -(\theta_1 - \theta_2) + 3\sqrt{2} \end{bmatrix} \quad (25)$$

where $\theta_1$ and $\theta_2$ are parameter values of two independent standard normal random variables $\Theta_1 \sim N(0,1)$ and $\Theta_2 \sim N(0,1)$, respectively.

Figure 3 shows how the support points are adaptively selected to construct the surrogate in S4IS. They are scattering over the all four failure regions especially the four failure boundaries. As a result, as can be seen from Figure 3, the original limit-state function can be well approximated by the surrogate.

Table 1 compares the results from S4IS with those from the crude MCS, FORM and AK-IS [18]. The failure probability from MCS $\hat{P}_{F,MCS}$ is considered as the ground truth of the problem and the relative error $\varepsilon_r$ is calculated by:

$$\varepsilon_r = \frac{\left|\hat{P}_{F,MCS} - \hat{P}_F\right|}{\hat{P}_{F,MCS}} \qquad (26)$$

where $\hat{P}_F$ is the failure probability obtained from the reliability analysis by other methods. Due to the nonlinearity of the limit-state function and the multiple failure regions as show in Figure 3, the failure probability from FORM is inaccurate with $\varepsilon_r = 69.8\%$. No improvement is observed in AK-IS as the importance sampling in the second stage is based on the FORM results in the first stage. For S4IS, an estimator with a large $CoV$ was obtained in Stage 1 based on the coarse surrogate; the estimator was improved significantly in Stage 2 resulting in $\varepsilon_r = 0.5\%$ and $CoV = 1.06\%$. Also, S4IS is more efficient compared with AK-IS, requiring only a total number of 60.6 evaluations of the performance function at the end of Stage 2.

It should be pointed out that for system reliability problems, we suggest to use the multi-output surrogate or multiple surrogates to approximate all component performance functions in the system then call the min (for series systems) or max (for parallel systems) function, rather than approximating the aggregated performance function only. The reason is that the min or max function of simple component performance functions may become too complex to be approximated accurately, leading to a large error of the final failure probability estimator.

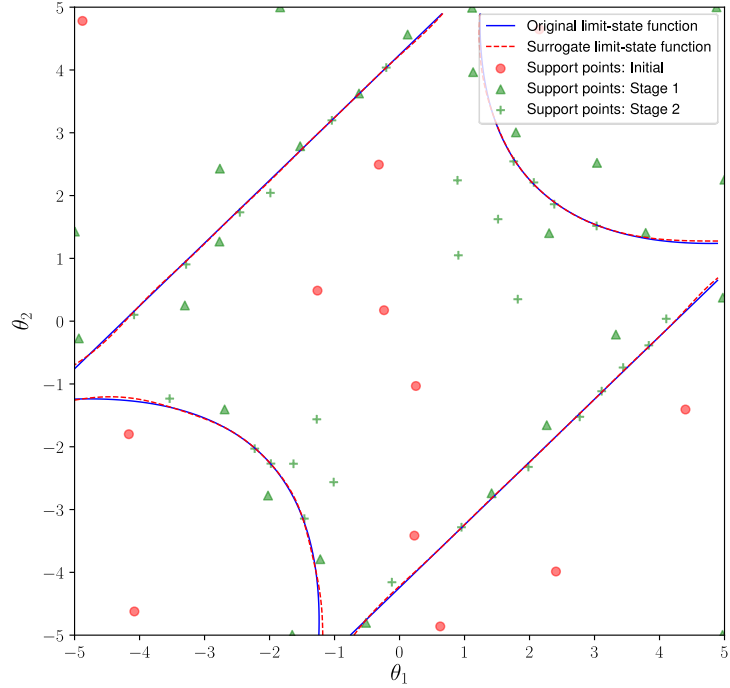

**Figure 3: Adaptively selected support points and the limit-state function using S4IS in Example 1**

**Table 1: Comparison of the results using different methods for Example 1**

| Methods | MCS | FORM | AK-IS | S4IS Stage 1 | S4IS Stage 2 |
|---|---|---|---|---|---|
| $\hat{P}_F$ | $4.460 \times 10^{-3}$ | $1.348 \times 10^{-3}$ | $1.179 \times 10^{-3}$ | $5.222 \times 10^{-3}$ | $4.483 \times 10^{-3}$ |
| $\varepsilon_r$ | -- | 69.8% | 73.6% | 17.1% | 0.5% |
| $CoV$ | 1.0% | -- | <5.0% | 13.7% | <5.0% |
| $N_{eval}$ | $10^6$ | 12 | 71.1 | 35.7 | 60.6 |

## 4.2 Example 2: a nonlinear oscillator [18, 25, 26]

Example 2 investigates the effect of the highly nonlinear limit-state function on the performance of S4IS. The oscillator of interest is shown in Figure 4 and the statistical properties

of its six random variables are summarized in Table 2. The performance function can be expressed as:

$$g(\mathbf{\theta}) = 3r - \left| \frac{2F_1}{m\omega_0^2} \sin\left(\frac{\omega_0 t_1}{2}\right) \right| \tag{27}$$

where $\mathbf{\theta} = [c_1, c_2, m, r, t_1, F_1]^T$ and $\omega_0 = \sqrt{\frac{c_1+c_2}{m}}$.

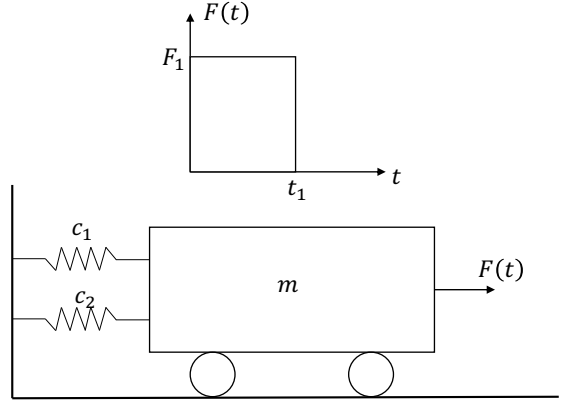

**Figure 4: Schematic diagram of the oscillator in Example 2**

**Table 2: Statistical characteristics of random variables in Example 2**

| Random variable | Distribution | Mean | Standard deviation |
|---|---|---|---|
| $c_1$ | Normal | 1 | 0.1 |
| $c_2$ | Normal | 0.1 | 0.01 |
| $m$ | Normal | 1 | 0.05 |
| $r$ | Normal | 0.5 | 0.05 |
| $t_1$ | Normal | 1 | 0.2 |
| $F_1$ | Normal | 1 | 0.2 |

The results of the reliability analysis using different methods are reported in Table 3. While the relative error of the estimated failure probability from FORM is large, both AK-IS and S4IS results in accurate estimators with $CoV < 5.0\%$ and $\varepsilon_r < 1.0\%$. Furthermore, S4IS only requires a total number of 53.3 calls to the performance function, being about half of the number required by AK-IS.

**Table 3: Comparison of the results using different methods for Example 2**

| Methods | MCS | FORM | AK-IS | S4IS |
|---|---|---|---|---|

|  |  |  |  | Stage 1 | Stage 2 |
|---|---|---|---|---|---|
| $\hat{P}_F$ | 0.02857 | 0.03116 | 0.02863 | 0.03347 | 0.02830 |
| $\varepsilon_r$ | -- | 9.1% | 0.2% | 17.2% | 0.9% |
| $CoV$ | 0.6% | -- | <5.0% | 40% | <5.0% |
| $N_{eval}$ | $10^6$ | 39 | 91.4 | 34.2 | 53.3 |

### 4.3 Example 3: a multimodal nonlinear function [9]

This example involves a highly nonlinear and multimodal performance function defined by:

$$g(\boldsymbol{\theta}) = -\frac{(\theta_1^2 + 4)(\theta_2 - 1)}{20} + \sin\left(\frac{5\theta_1}{2}\right) + 2 \tag{28}$$

Two independent random variables follow the normal distribution $\Theta_1 \sim N(1.5,1)$ and $\Theta_2 \sim N(2.5,1)$.

As can been seen from Figure 1, as a whole, the original limit-state function can be approximated by the surrogate accurately with the multiple modals identified. What is interesting to observe in this figure is that the surrogate has a very high accuracy over the important region where the density value of the distribution of random variables is large. It is a natural result of the adaptive infill strategy as in S4IS the second stage zooms into the important region and produces support points therein.

Table 3 summarizes the results by using different reliability analysis methods. Both AK-IS and S4IS can achieve the small relative error and variance in the failure probability estimation. In this example, S4IS outperforms greatly AK-IS in terms of efficiency as it takes hundreds of iterations for the optimization procedure in the first stage of AK-IS to find the most probable point (MPP).

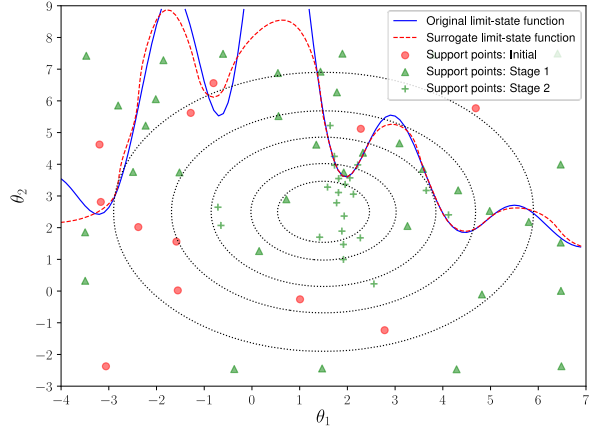

**Figure 5:** Adaptively selected support points and the limit-state function using S4IS in Example 3. The dotted lines are the contour lines of the joint probability density function.

**Table 4: Comparison of the results using different methods for Example 3**

| Methods | MCS | FORM | AK-IS | S4IS | |
| --- | --- | --- | --- | --- | --- |
| | | | | Stage 1 | Stage 2 |
| $\hat{P}_F$ | 0.03130 | 0.1182 | 0.03123 | 0.02049 | 0.03078 |
| $\varepsilon_r$ | -- | 277.6% | 0.2% | 34.5% | 1.7% |
| $CoV$ | 0.3% | -- | <5.0% | 61.5% | <5.0% |
| $N_{eval}$ | $10^6$ | 695 | 985.9 | 46.2 | 71.4 |

### 4.4 Example 4: the influence of reliability levels [11, 27]

The purpose of this example is to investigate the influence of different reliability levels on the performance of S4IS, particularly when the reliability level is very high (small failure probability). The performance function reads:

$$g(\boldsymbol{\theta}) = \min \left\{ \begin{array}{c} c - 1 - \theta_2 + \exp\left(-\frac{\theta_1^2}{10}\right) + \left(\frac{\theta_1}{5}\right)^4 \\ \frac{c^2}{2} - \theta_1 \theta_2 \end{array} \right\} \quad (29)$$

Two independent random variables in the performance function follow the standard normal distribution. By increasing the constant $c$, the reliability level can be increased accordingly.

To investigate the influence of different reliability levels, three cases ($c = 3$, $c = 4$ and $c = 5$) are discussed. Comparative results are reported from Table 5 to Table 7. S4IS performs consistently well for different reliability levels, in which the coarse estimator $\hat{P}_F$ in Stage 1 can be improved significantly in Stage 2. In this example, increasing the reliability level from $\hat{P}_F = 3.470 \times 10^{-3}$ to $\hat{P}_F = 9.485 \times 10^{-7}$ only add a little computational effort (from $N_{eval} = 72.8$ to $N_{eval} = 118.6$) to achieve the high accuracy. Again, FORM and AK-IS tend to be mistaken by the only MPP they can find, resulting in the inaccurate estimators.

Table 5: Comparison of the results using different methods for Example 4 ($c = 3$)

| Methods | MCS | FORM | AK-IS | S4IS | |
|---|---|---|---|---|---|
| | | | | Stage 1 | Stage 2 |
| $\hat{P}_F$ | $3.470 \times 10^{-3}$ | $1.350 \times 10^{-3}$ | $1.462 \times 10^{-3}$ | $3.941 \times 10^{-3}$ | $3.531 \times 10^{-3}$ |
| $\varepsilon_r$ | -- | 61.1% | 57.9% | 13.6% | 1.8% |
| $CoV$ | 1.3% | -- | <5.0% | 35.7% | <5.0% |
| $N_{eval}$ | $10^6$ | 7 | 97.6 | 44.9 | 72.8 |

Table 6: Comparison of the results using different methods for Example 4 ($c = 4$)

| Methods | MCS | FORM | AK-IS | S4IS | |
|---|---|---|---|---|---|
| | | | | Stage 1 | Stage 2 |
| $\hat{P}_F$ | $9.172 \times 10^{-5}$ | $3.167 \times 10^{-5}$ | $4.509 \times 10^{-5}$ | $9.286 \times 10^{-5}$ | $9.120 \times 10^{-5}$ |
| $\varepsilon_r$ | -- | 65.5% | 50.8% | 1.2% | 0.6% |
| $CoV$ | 4.8% | -- | <5.0% | 9.3% | <5.0% |
| $N_{eval}$ | $4 \times 10^6$ | 7 | 110.3 | 54.4 | 83.2 |

Table 7: Comparison of the results using different methods for Example 4 ($c = 5$)

| Methods | MCS | FORM | AK-IS | S4IS | |
|---|---|---|---|---|---|
| | | | | Stage 1 | Stage 2 |
| $\hat{P}_F$ | $9.485 \times 10^{-7}$ | $2.867 \times 10^{-7}$ | $2.277 \times 10^{-7}$ | $6.060 \times 10^{-7}$ | $9.035 \times 10^{-7}$ |
| $\varepsilon_r$ | -- | 69.8% | 76.0% | 36.1% | 4.7% |
| $CoV$ | 4.9% | -- | <5.0% | 9.1% | <5.0% |
| $N_{eval}$ | $4 \times 10^8$ | 7 | 92.4 | 63.4 | 118.6 |

## 4.5 Example 5: the influence of dimensionality [11, 28]

The last example is characterized by an analytical performance function where the number of random variables $d$ can be changed without altering the reliability level a lot. The performance function reads as follows:

$$g(\boldsymbol{\theta}) = \left(d + 6\sqrt{d}\right) - \sum_{i=1}^{d} \theta_i \qquad (30)$$

This example involves $d$ independent lognormal random variables with the mean values 1 and standard deviations 2.

Three cases ($d = 2$, $d = 10$ and $d = 50$) are investigated by using different methods for this example. The results are summarized from Table 8 to Table 10. In S4IS, for moderately high dimensional cases ($d = 10$ and $d = 50$)), the combination of the search of multiple MPPs and FORM [1] is adopted in Stage 1 as discussed in Section 3.1.4. It is shown that on this example, the total required number of evaluations $N_{eval}$ in S4RS increases slowly with the number of random variables. For $d = 50$, less than two hundred evaluations are required, which is generally affordable in the setting of the simulation of the real-world systems.

The results using other reliability analysis methods are compared to S4IS. For FORM, though the performance space in the original space is linear, it becomes nonlinear when the original random variables are transformed into the standard normal ones. As it can be seen, FORM fails when the dimension is high since the nonlinearity of the limit-state function in the U-space. AK-IS can produce the accurate estimators of the failure probability for this example even when the dimension is high, but AK-IS requires much more calls of the performance function than S4IS (for $d = 50$, $N_{eval} = 1845.2$ in AK-IS). It is not surprising as the stopping criterion in AK-IS rely on the convergence of the minimum U-function value over the all IS samples while S4IS checks the convergence of the estimator directly.

Table 8: Comparison of the results using different methods for Example 5 ($d = 2$)

| Methods | MCS | FORM | AK-IS | S4IS | |
|---|---|---|---|---|---|
| | | | | Stage 1 | Stage 2 |
| $\hat{P}_F$ | $4.926 \times 10^{-3}$ | $3.844 \times 10^{-3}$ | $4.928 \times 10^{-3}$ | $4.936 \times 10^{-3}$ | $4.921 \times 10^{-3}$ |
| $\varepsilon_r$ | -- | 22.0% | 0.04% | 0.2% | 0.1% |
| $CoV$ | 1.6% | -- | <5.0% | 15.4% | <5.0% |
| $N_{eval}$ | $10^6$ | 20 | 59.0 | 16.4 | 23.9 |

Table 9: Comparison of the results using different methods for Example 5 ($d = 10$)

| Methods | MCS | FORM | AK-IS | S4IS | |
|---|---|---|---|---|---|
| | | | | Stage 1 | Stage 2 |
| $\hat{P}_F$ | $2.744 \times 10^{-3}$ | $1.003 \times 10^{-3}$ | $2.711 \times 10^{-3}$ | $1.523 \times 10^{-3}$ | $2.739 \times 10^{-3}$ |
| $\varepsilon_r$ | -- | 63.4% | 1.2% | 44.5% | 0.2% |
| $CoV$ | 1.5% | -- | <5.0% | -- | <5.0% |
| $N_{eval}$ | $10^6$ | 35 | 678.2 | 38 | 48.6 |

Table 10: Comparison of the results using different methods for Example 5 ($d = 50$)

| Methods | MCS | FORM | AK-IS | S4IS | |
|---|---|---|---|---|---|
| | | | | Stage 1 | Stage 2 |
| $\hat{P}_F$ | $1.934 \times 10^{-3}$ | $1.541 \times 10^{-4}$ | $1.903 \times 10^{-3}$ | $2.318 \times 10^{-4}$ | $1.915 \times 10^{-3}$ |
| $\varepsilon_r$ | -- | 92.0% | 1.6% | 88.0% | 1.0% |
| $CoV$ | 1.3% | -- | <5.0% | -- | <5.0% |
| $N_{eval}$ | $10^6$ | 155 | 1845.2 | 157 | 168.6 |

## 5 Conclusions

This paper proposed a new framework for surrogate-aided reliability analysis called Surrogates for Importance Sampling (S4IS). This framework is very efficient in building the surrogate by selecting support points adaptively. The basic idea is to use different learning functions and candidate support points in the two stages such that the exploration and exploitation capability of the selected support points can be dynamically balanced. After gaining the information about the failure regions in the first stage, the second stage zooms into the important regions to improve the accuracy of the failure probability estimator. Multiple failure regions can be identified in the first stage via the clustering of the failure samples or the search of most probable points directly.

The proposed S4IS has been validated by five illustrative examples for different types of reliability analysis problems, which are featured by system reliability, nonlinear limit-state functions, small failure probability and moderately high dimensionality. Compared to other reliability analysis methods, S4IS performs well in all cases meanwhile it requires a small number of evaluations of the performance function for the failure probability estimator to achieve the small relative error and coefficient of variation. For the example studied, S4IS is robust to the number of random variables up to 50.

Surrogate-aided reliability analysis opens promising ways to tackle the Achilles' Heel of reliability analysis, which is the computational burden especially expensive-to-evaluate performance functions and stochastic-sampling-based methods are involved. Despite the specific surrogate Gaussian Process used in this paper, the proposed framework is applicable to other types of surrogate. In future research, the effect of different types of surrogates on S4IS should be investigated.